# Review of Ultrafast Spectroscopy Studies of Valley Carrier Dynamics in Two Dimensional Semiconducting Transition Metal Dichalcogenides[*]

**Dong Sun (孙栋)**[1,2,†], **Jia-Wei Lai (赖佳伟)**[1], **Jun-Chao Ma (马骏超)**[1], **Qin-Sheng Wang (王钦生)**[1], **Jing Liu (刘晶)**[3]

[1] *International Center for Quantum Materials, School of Physics, Peking University, Beijing 100871, P. R. China*
[2] *Collaborative Innovation Center of Quantum Matter, Beijing 100871, P. R. China*
[3] *State Key Laboratory of Precision Measurement Technology and Instruments, School of Precision Instruments and Optoelectronic Engineering, Tianjin University, Tianjin, People's Republic of China*

The two-dimensional layered transition metal dichalcogenides provide new opportunities in future valley based information processing and also provide ideal platform to study excitonic effects. At the center of various device physics toward their possible electronic and optoelectronic applications is understanding the dynamical evolution of various many particle electronic states, especially exciton which dominates the optoelectronic response of TMDs, under the novel context of valley degree of freedom. Here, we provide a brief review of experimental advances in using helicity resolved ultrafast spectroscopy, especially ultrafast pump-probe spectroscopy, to study the dynamical evolution of valley related many particle electronic states in semiconducting monolayer transitional metal dichalcogenides.

## 1. Introduction

Two dimensional (2D) semiconducting transition metal dichalcogenides (TMDCs) have attracted enormous research interests in recent years due to their optically addressable valley degree of freedom [1-4]. This novel degree of freedom provides additional opportunity beyond charge and spin that have been routinely explored in conventional devices. On the other hand, due to strong quantum confinement and reduced screening in atomically thin 2D limit, the binding energies of many particles electronic states, such as exciton, trion and biexciton, are extremely large comparing to conventional semiconductor and nanostructures [5-11]. Despite early experimental milestones have been achieved in demonstrating both the injection and detection of valley polarized exciton states in monolayer TMDCs using photoluminescence (PL) technique[2-4] and valley Hall effect[12] even before the understanding of the detail valley carrier dynamics, these successful conceptual demonstration turns out to be lucky successes as the details of valley carrier dynamics are gradually revealed by many ultrafast spectroscopy works later on. On the other hand, the valley carrier dynamics, namely the valley-depolarization mechanism and the timescale of valley lifetime, are key device physics that determines device operations and information storages [13-17].

[*] Project supported by the National Basic Research Program of China (973 Grant Nos: 2012CB921300 and 2014CB920900), National Key Research and Development Program of China (Grant No: 2016YFA0300802), the National Natural Science Foundation of China (NSFC Grant Nos: 11274015, 11674013 and 21405109) and the Recruitment Program of Global Experts.
[†] Corresponding author. E-mail: sundong@pku.edu.cn

Ultrafast pump-probe spectroscopy is a powerful experimental tool to study various carrier dynamics in solids. In a pump-probe experiment, the samples are excited with a pump pulse and then the evolution of carriers' properties of the system is probed at different delay times respect to the pump pulse. During the past few years, a number of ultrafast time-resolved spectroscopy techniques have been applied to study various TMDs. These include time-resolved photoluminescence [18-25], transient absorption spectroscopy [20, 26-37] and transient Kerr rotation technique [38-42]. These ultrafast spectroscopy techniques, combining with helicity resolution of light, are versatile and powerful tools in addressing the valley degree of freedom of carriers taking advantage of the valley contrast circular dichroism of monolayer TMDs. However, most interpretations of ultrafast spectroscopy results in the literature are still under debates and some of them are even in confliction with others due to the complication of data interpretation, therefore a coherent self-consistent diagram of existing reported experimental work is still highly desired, although there are already a number of reviews regarding carrier dynamics in TMDs.[43-50] In this work, we mainly review various helicity resolved ultrafast spectroscopy measurements regarding valley carrier dynamics in TMDs. To be more focused on the valley degree of freedom, ultrafast spectroscopy works without helicity resolution are mostly not included in the discussion. In the following, we will first briefly introduce valley contrast circular dichroism and various many particle electronic states in monolayer TMDs; then we'll discuss various helicity resolved ultrafast spectroscopy measurements results on monolayer TMDs, including time resolved photoluminescence, ultrafast transient absorption measurements and ultrafast Kerr rotation measurements. The major goal of this review is reinvestigating various interpretations of different helicity resolved ultrafast spectroscopy measurements and try to give uniform interpretation regarding valley related carrier dynamics in monolayer TMDs from experimental results reported so far.

2. **Valley Exciton in 2D Transitional Metal Dichalcogenides**

2H phase transitional metal dichalcogenides have a hexagonal lattice structure. Taking monolayer $MoS_2$ as example, it consists of a single layer of molybdenum atoms sandwiched between two layers of sulphur atoms in a trigonal prismatic structure (Fig. 1(a)). As the two triangle sublattices are occupied by different atoms, the inversion symmetry is broken, which gives rise to a valley-contrasting optical selection rule, where the inter-band transition in the vicinity of the K (K') point couple exclusively to right (left) circularly polarized light.[1, 4] Furthermore, spin–orbit interactions split the valence bands by 160 meV in MoS$_2$[1] (400 meV in WS$_2$[51, 52] ), the spin projection along the c-axis of the crystal is well defined and the two bands are of spin down (E↓) and spin up (E↑) in character. The splitting of the conduction band, however, is tiny, giving almost degenerated energy band at conduction band minimum. The band-edge transitions in monolayer TMDs are strongly modified by electron–hole (e–h) interactions due to reduced screening effect in monolayer crystal as discussed in next session, which gives rise to A and B excitons. The selection rules, however, carry over to the excitons as illustrated in Fig. 1(b).

The direct gap of monolayer TMDs[53, 54] at K point great facilitate the experimental breakthrough in manipulation of valley pseudospin, especially the exciton emission of monolayer TMDs falls into experimentally matured visible and near-IR range. The valley contrast circular dichroism was first demonstrated using polarization resolved PL[2-4]. In these experiments, circularly polarized light selectively excites an A exciton in one valley and the resulting PL is found to be strongly polarized with the same circular polarization as the excitation light (Fig. 1(c)). This result provides direct evidence that the valley polarization of the photo-excited electrons and holes is largely retained during exciton formation, relaxation and radiative recombination processes. Despite the valley contrast selection rule is under expectation, the near unity PL polarization observed on MoS$_2$ is surprising. In these polarized PL measurements, the measured PL polarization P=1/(1+2$\tau_r$/$\tau_v$), where $\tau_r$ is the radiative life time of exciton and $\tau_v$ is the intervalley scattering time. The near unity PL polarization observed in a steady state PL polarization measurement[2] is a result of very short radiative life time of exciton in TMDs due to extremely large exciton binding energy. Furthermore, Hanle effect is used to test whether the polarized PL is due to the polarization of the valley instead of spin, as a transverse magnetic field will cause spin to process, but not valley[3].

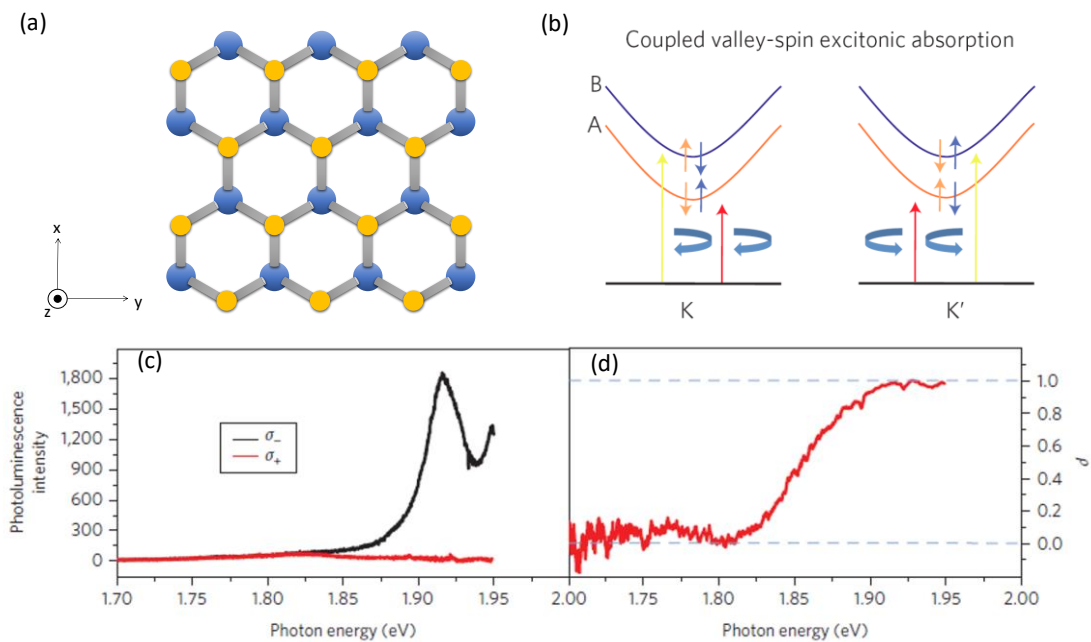

**Fig. 1.** (a) 2D hexagonal lattice of monolayer TMDs of x-y plane. It has structure that is lacking of inversion symmetry. (b)Optical selection rules in monolayer MoS$_2$ for the A and B exciton states at

K (K') valley coupled with left (right) circularly polarized light. (c) Left (σ⁻) and right (σ⁺) circularly polarized resolved photoluminescence spectra for monolayer MoS$_2$ on h-BN substrates. (d) Corresponding photoluminescence helicity ρ. Panels (b) (c) and (d) are reproduced from Ref. [2]

### 2.1. Many-particle Electronic States

Coulomb interactions can lead to the formation of stable many-particle electronic states, such as excitons. Exciton is a bound electron and hole pair with an energy spectrum similar to that of a hydrogen atom. In doped semiconductors, a neutral exciton can bind to an extra electron or hole to form a charged exciton (trion). On the other hand, two excitons can be bound by residual Coulomb fields to form a biexciton state. Monolayer TMDs can host all these many-particle electronic states, as enhanced many-body Coulomb interactions emerge in atomically thin materials due to the extreme quantum confinement and reduced screening yields. As monolayer TMDs are direct gap semiconductors, these excitonic quasi-particles appear as pronounced resonances in the optical response and can dominate absorption and emission spectra.

Here we take MoS$_2$ as example to discuss the many-particle states in monolayer TMDs. Figure 2(a) shows the simplified band structure of MoS$_2$, A and B are the direct-gap transitions from the two spin-orbital spitted valence bands[53, 54]. Absorbance of mono layered MoS$_2$[55] shown in Fig. 2(b) indicates three absorption peaks which are attributed to three different excitonic transitions which are usually denoted as A (1.88 eV), B (2.06 eV) and C (3.1 eV) excitons, among which A and B correspond to transitions of spin-orbital split bands and C corresponds to the transitions from the almost parallel part of conduction and valence bands as marked in Fig. 2(a), but all are modified with coulomb interaction of electron and hole. The interaction between light and monolayer TMDs is very prominent comparing to the 2.3% absorption of graphene: the absorbance is around 10% at photon energy in resonance with A/B excitons and keep climbing with photon energy, at 3.1 eV, it's over 20% due to band nesting effect of C band.[55-58] Due to the extreme quantum confinement and reduced screening yields, the exciton binding energy in monolayer TMDs ranges from 0.5 to 0.9 eV according to theoretical calculations[59-61] and optical/scanning tunneling spectroscopy measurement.[5-8]

When there are excess electron or hole available, a charged exciton (trion) can form as shown in Fig. 2(c). Trion binding energy of 20 meV and 30 meV are reported on MoS$_2$ and MoSe$_2$ respectively [9, 10]. As the doping can be tuned with a gate in 2D monolayer, the evolution of a trion state, from positive trion to neutral and then negative trion is doping dependent as measured in gate dependent PL and broadband reflection measurement shown in Fig 2(d). With high exciton density, two excitons can also bind together to form energy stable biexciton state, as the coulomb interaction can further reduced the total energy if the two excitons binds together (Fig 2(e)). As demonstrated experimentally by You et al.[11], under high intensity pulse excition (~12 μJ/cm$^2$), the biexciton is identified as a sharply defined state in photoluminescence (Fig. 2(f)). Its binding energy of 52meV is more than an order of magnitude greater than that found in conventional quantum-well structures.[11] Due to the large binding energy comparing to conventional quasi-2D semiconductors, most of these excitonic states can survive at room temperature in monolayer TMDs. These have important consequences for optical absorption spectrum of monolayer TMDs as significant oscillation strength is transferred from band-to-band transition to resonant absorption of excitonic states.[62]

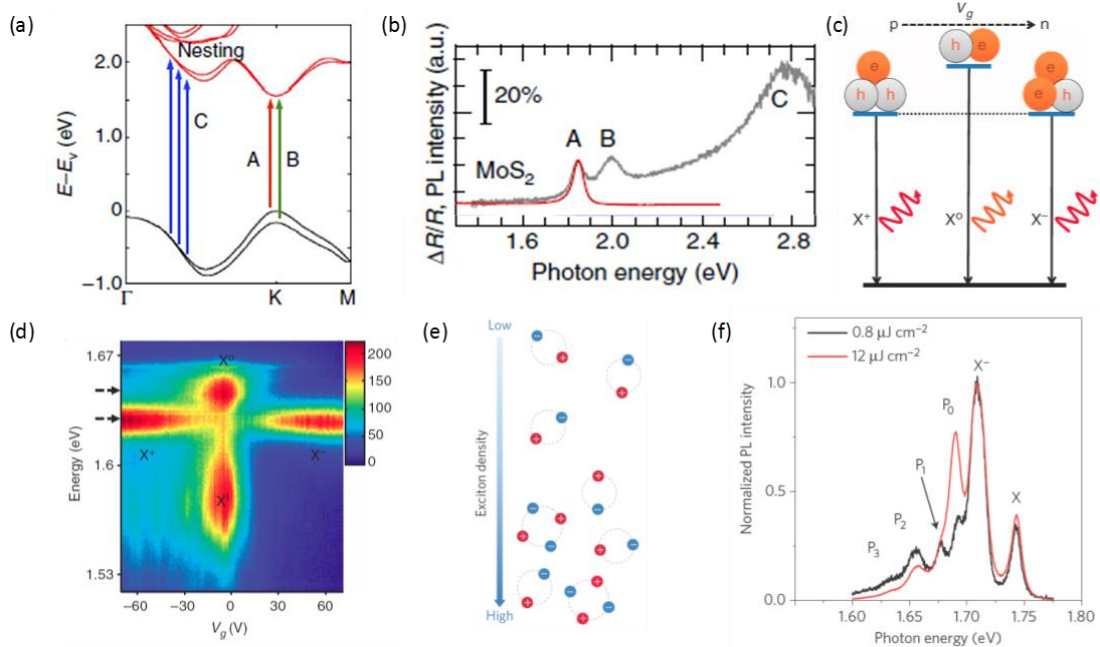

**Fig. 2.** (a) The band structure of monolayer $MoS_2$. The arrows indicate the transitions in A, B and the band nesting C. (b) PL spectra (red) from excitation at the C peak and differential reflectance spectra (grey) of monolayer $MoS_2$ flake on quartz substrate. The scale bar indicates 20% absorption based on the differential reflectance spectra. The PL intensity is normalized by the A exciton peak of the differential reflectance spectra. (c)Gate-dependent trion and exciton quasi-particles and transitions. (d) PL of $MoSe_2$ plotted as a function of back-gate voltage. Neutral and impurity-trapped excitons are observed mostly at near zero doping. Charge excitons dominate the spectrum with large electron (hole) doping, negatively (positively) charged excitons. (e)Schematic diagram of biexcitons formation from excitons as four-body quasi-particles with increasing exciton density. (f)PL spectra at 50 K for pulsed excitation under applied fluence of 0.8 μJ cm$^{-2}$ and 12 μJ cm$^{-2}$. The spectra are normalized to yield the same emission strength for the neutral exciton. Panels (a) and (b) are reproduced from Ref. [63]; Panels (c) and (d) are reproduced from Ref. [10]; Panels (e) and (f) are reproduced from Ref. [11].

### 3. Ultrafast Spectroscopy Studies of Coupled Spin-Valley Dynamics

In the following, we will discuss the progress of studying coupled spin-valley dynamics in TMDs mainly through helicity resolved ultrafast pump-probe spectroscopy techniques. The discussions of helicity resolved TRPL results are also included, the TRPL results are strongly correlated to the coupled spin-valley dynamics. On the other hand, pump-probe spectroscopy work without helicity resolution is not included except the measurement of band renormalization effect, as it's generally not related to the spin or valley dynamics.

## 3.1. Time Resolved Photoluminescence

Although TRPL is not considered as a pump-probe technique, it provides clear and straight forward life-time measurement of photon emission from radiative recombination of photo excited carriers. Combining with helicity resolution of pump excitation and emitted photon detection, helicity resolved TRPL measurements reveal the spin-valley coupled polarization dynamics within the interval between exciton/trion initial excitation and radiative recombination through emission photon. Before the optical addressable valley degree of freedom of TMDs was discovered in 2012, T. Korn et al. has already studied the radiative lifetime of exciton in monolayer $MoS_2$ using TRPL.[18] As shown in Fig. 3(a), the radiative recombination lifetime is found to be 4.5 picosecond at 4 K. As temperature increases, a bi-exponential decay with a long-lived component is developed, at room temperature, the long-lived component is increased to about 70 ps monotonously with temperature. This long-lived PL at high temperature is attributed to the exciton scattered out of light cone through exciton-phonon scattering, after the scattering, the excitons have to reduce their momenta again before they may recombine by radiation.

Helicity resolved TRPL measurement of exciton radiation were performed by different research groups on mechanically exfoliated monolayer $MoS_2$[19], $WSe_2$[20, 21] and $MoSe_2$[22] respectively. In $WSe_2$ and $MoSe_2$ samples, the trion emission dynamics are also studied.[20-22] Except the measurements on $MoSe_2$ sample showing very low circular polarization for both exciton and trion states[22], which remains a mystery and under debate in the community; measurements on monolayer $MoS_2$ and $WSe_2$ reveal clear dynamical evolution of the helicity of the emitted photon through recombination of valley polarized exciton or trion state with few picosecond resolution. Figure 3(b) shows helicity resolved TRPL measurement on $WSe_2$ sample at 4 K.[23] The exciton emission time cannot be resolved which is within 4 ps temporal resolution, indicating the radiative recombination time of exciton is faster than 4 ps. The PL emission with orders of magnitude weaker at later times can be fitted by exponential decay with a characteristic time of 33±5 ps, this is attributed by the authors to the exciton localized at fluctuations of the crystal potential.[24, 25] The trion PL emission can be fitted by a bi-exponential decay with the initial decay of 18±2 ps, which is longer than the exciton. The longer decay time is about 30±3ps, similar to that of exciton.

Figure 3(c) shows the helicity resolved TRPL measurement on exciton at 4 K[23], as the exciton emission time is below the time resolution, the valley polarization only exist during the laser excitation region marked by the shade area determined by the temporal resolution. The initial trion emission is much longer and the time evolution of the valley polarization can be accessed. As shown in Fig. 3(d), the trion polarization decays with 12 ps from 50% down to 20%, followed by a second decay with a characteristic time of the order of 1 ns. The first decay can be ascribed to the intervalley exchange interaction similar to that of exciton state, after which, the trion polarization decay dynamics evolves differently, it gets slower as the majority of exciton has recombined. The second polarization decay is slower as the single particle spin flips are needed in addition to the intervalley scattering for trion state. The much longer coupled spin-valley polarization life time of trion, comparing to neutral exciton state is consistent with the time resolved Kerr measurement on valley-spin polarization of the trion and resident electron/hole in TMDs, which will be discussed in Session 3.3. In a parallel measurement on WSe$_2$ by Yan T et al.[21], the long-lived component of trion polarization is not observed, this is probably due to the elevated measurement temperature (100K in their experiment).

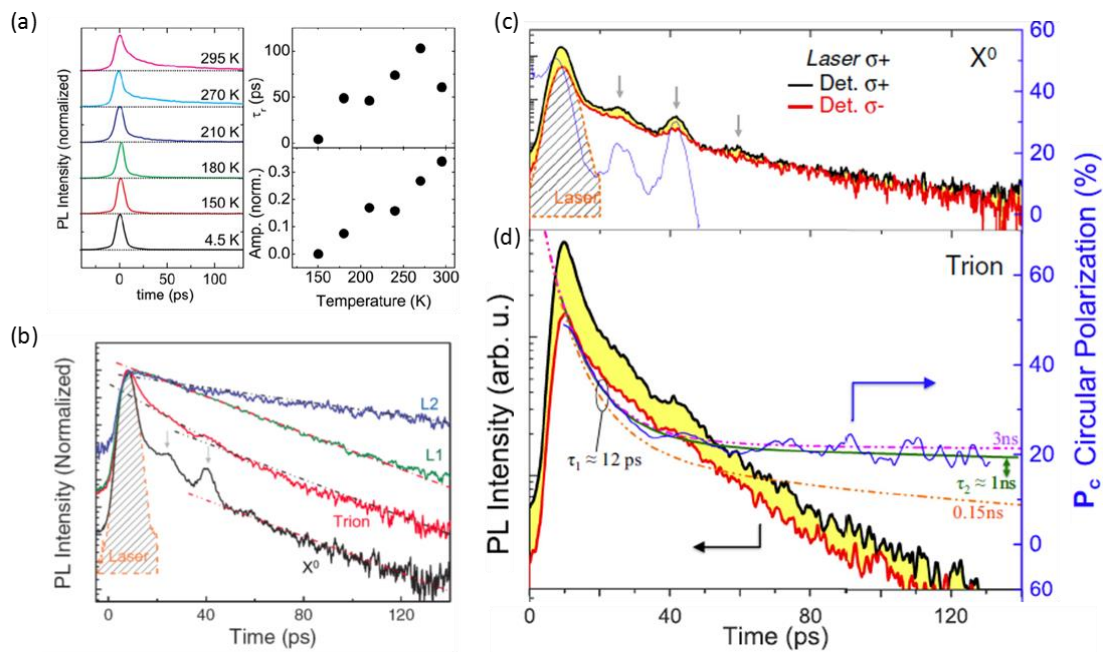

**Fig. 3.** (a)Left panel: normalized TRPL traces measured on the monolayer part of the MoS$_2$ flake at different temperatures from 4.5 K to 295 K. Right panel: plots of slow decay time constant (upper panel) of photo carrier recombination ($\tau_r$) and amplitude of the TRPL traces (lower panel) as a function of temperature. (b)Normalized time-resolved PL dynamics at 4 K in log scale. The lines are all fitted by bi-exponential function, however, the initial time is too short to be fitted for X$^0$, L1 and L2. The decay times are 33±5ps for the second decay time of X$^0$, 32±2ps for L1, 80±6ps for L2. For trion, the short time constant is 18±2ps and the long decay time constant is 30±3ps. The small peaks marked by arrows superimposed on the X$^0$ and trion come from laser reflections. (c)(d)TRPL at 4 K with σ+ laser polarization excitation Left axis is X$^0$ (c) and trion (d) PL emission (in log scale) with co-polarized (black lines) and cross-polarized (red lines) detection geometry with respect to the excitation laser as a function of time, and right axis is circular polarization degree of the PL emission. (c)Arrows show the periodic signal of laser reflections as the PL intensity decays. (d)Solid green line shows the polarization reproduced by a bi-exponential decay, using a fast decay time of $\tau_1$=12 ps and a long decay time $\tau_2$=1 ns. Dotted orange line shows lower bounds for $\tau_2$=150 ps and dotted purple line shows upper bounds for $\tau_2$=3 ns. Panel (a) is reproduced from Ref. [18]; Panels (b) (c) and (d) are reproduced from Ref. [23].

### 3.2. Helicity Resolved Transient Transmission/Reflection Spectroscopy

Limited by the detection of radiative emission only, non-radiative evolution of carriers is "invisible" in a TRPL measurement. Alternatively these non-radiative process can be sensed in a transient absorption (TA) spectroscopy measurement in either transmission or reflection geometries as described in this session. In a helicity resolved transient absorption (HRTA) measurement scheme shown in Fig. 4(a), upon excitation of one valley with circularly polarized pump pulses, the measurement of transmission/reflection of the same circularly polarized probe (SCP) pulses at different time delays correlate to the decay of the pump excited valley polarization, whereas the measurement of opposite circularly polarized (OCP) pulses correlates to the rise of the population in the other valley. The difference between the TA signal of SCP and OCP probe thus correlates to the valley carrier distributions. The disadvantage of transient reflection/transmission measurement is the complexity of data interpretation, as the pump induced transient probe signal can be related to multiple effects and possibilities, and the quantitative determination of each effects is challenging and almost impossible. The HRTA measurements have been attempted on $MoS_2$ and $MoSe_2$ using fix photon energy probe around A exciton resonance by Wang et al.[26] and Nardeep et al.[27], however, the observed difference of transient signal between OCP and SCP only last several ps in $MoS_2$, which is mainly attributed to short exciton life time limited by defects trapping.

Furthermore, the same HRTA idea has been attempted with broadband probe on $MoS_2$ and $WS_2$[20, 28-30], which provides complete spectra evolutions around various excitons transitions. Figure 4(d) shows a typical transient absorption spectra of monolayer $MoS_2$ at 74K with various delays between pump and probe pulses obtained by Mia et al..[28] The difference between OCP and SCP, which correlates to exciton valley relaxation, only persists less than 15 ps in $MoS_2$ at 74 K. This decay time of the transient signal difference between OCP and SCP varies with samples. In a parallel experiment measured on $WS_2$ [30], this decay time increases to over 100 ps at 110 K. In both works, the electron-hole exchange interaction as illustrated in Fig. 4(b) and electron/hole spin relaxation via scattering through the spin degenerated T valley as illustrated in Fig. 4(c) are considered as possible dominated mechanism accounts for the intervalley scattering. As the pump excitation of carrier population in certain valley can induce very rich effects to the probe transmission/reflection signal, these effects generally fall into two categories: first, phase space filling effect induced by pump excited carriers and band renormalization. The first category includes ground state bleaching (GSB), which is a result of the Pauli blocking of probe transition due to pump excited carrier population in a band; photo induced absorption (PIE), which comes from new optical transition at probe photon energy due to the existence of pump excited state population; and stimulated emission (SE), which is a stimulated amplification effect when pump excited electron hole pair matches probe photon transition in the same valley. For the second category, both the bandgap and exciton binding energy is shifted due to pump excited carriers, which will be discussed in detail in the following session. Due to these complex natures of the HRTA measurements, the interpretations of intervalley scattering solely from transient absorption results in the early works are not complete. Time resolved Kerr measurements are usually serving as complementary information as discussed in later sessions.

Despite there are differences related to the nature of the samples in transient absorption studies, some common features are ubiquitously observed. First, there is always a bleaching of the lowest-energy excitonic features accompanied by red-shifted photo induced absorption sidebands as shown in Fig. 4(e). The interpretation to this feature is controversial in early works, ranging from carrier induced broadening[20], to exciton-bi exciton transition.[28-30] In the most recent HRTA measurements performed by Pogna et al.[55], it reports that this feature occurs for all A, B, C excitonic transitions irrespective to excitation photon energy. Combining with first principle modeling, they reveal that a transient bandgap renormalization caused by the presence of photo-excited carriers is primarily responsible for the observed features, this effect will be discussed in detail in the following session.

The second ubiquitous feature is a strong signal in the un-pumped K' valley arise immediately (within the experimental temporal resolution) after pumping the K valley. The population of K' valley can be either directly excited by pump excitation or through very rapid intervalley scattering within 100 fs temporal resolution of the measurement. Early work by Wang et al. proposed the carriers in the K' valley is populated directly by the pump excitation, a first principle simulation including disorder as shown in Fig. 4(e) indicates the helicity dependent valley selection rule is significantly released at non-resonant excitation conditions when significant disorder is present.[26] Later on, Mai et al. proposed the electron population created by the pump pulse in one valley spreads over both valleys due to the degeneracy of the conduction bands forming 'dark exciton" states.[28] Another interpretation by Sie et al. involved the creation of biexcitons with binding energies on the order of 40 meV. Recently Schmidt et al. further compares the HRTA measurement results and performed microscopic calculations based on semiconductor Bloch equations and finally nail down the answer to this question: they show that Coulomb induced band gap renormalization and impurity induced intervalley carrier transfer during the optical excitation mainly account for the prominent spectral features in both valleys immediately after pump excitation.[31] As illustrated in Fig. 4(f), the intrinsic intervalley scattering (labeled as Y mechanism) couples states at the K and K' valley with the same spin (B exciton in K valley and A exciton in K' valley for example), and the impurity assisted electron−hole exchange coupling (labeled as X mechanism), couples energetic resonant states in the K and K' valley. The origin of the Y coupling is an overlap of intraband orbital functions, whereas the Coulomb exchange X coupling stems from an overlap of interband orbital functions. It couples resonant states at different valleys (A−A and B−B) and induces a microscopic polarization in the unpumped valley resulting in a carrier transfer. Schmidt et al. further shows these two effects together accounts for the appearance of a strong immediate signal in the OCP observed in HRTA measurements.

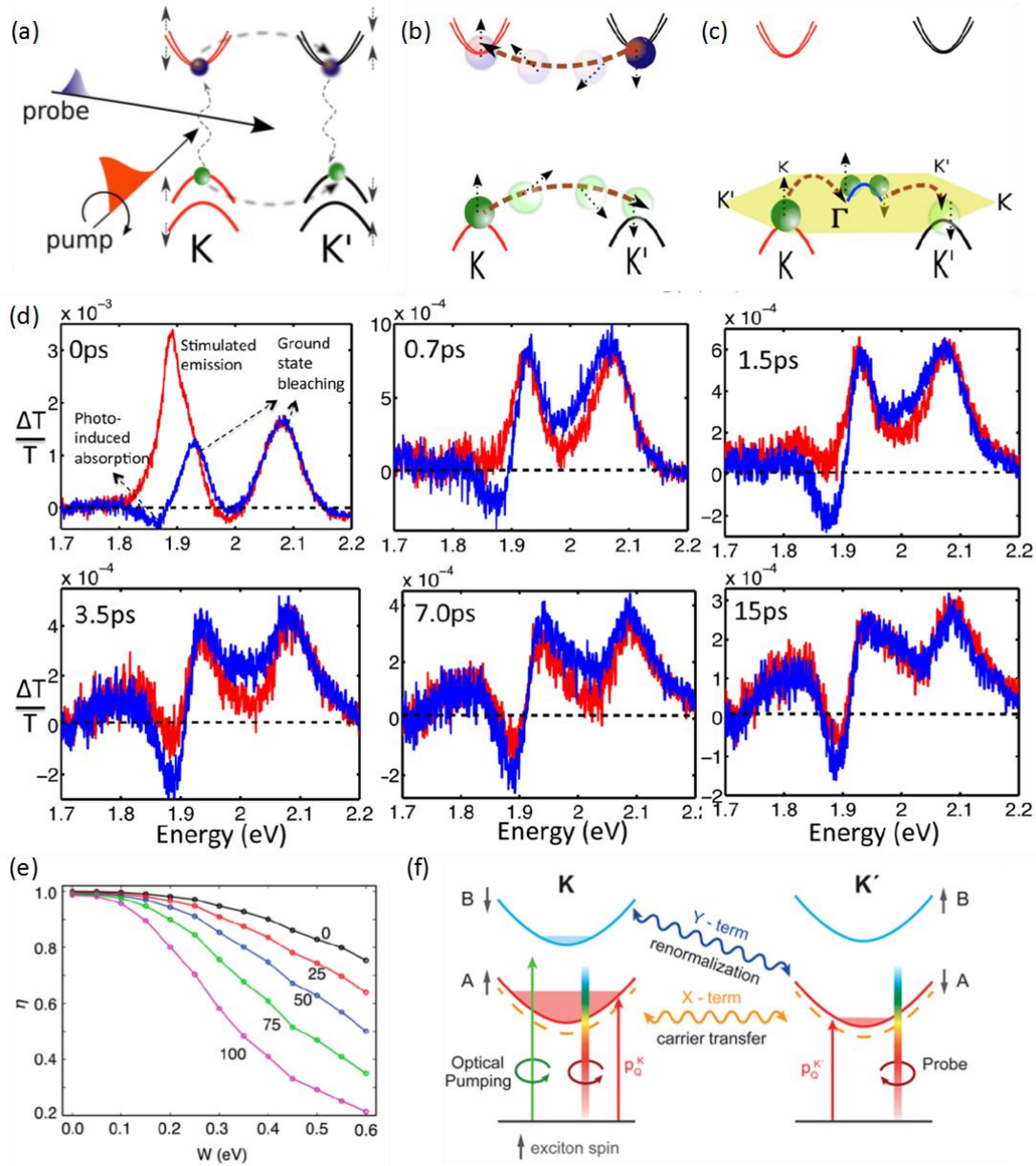

**Fig. 4.** (a)Pump beam with circular polarization excites K valley only, probe beam with the same or the opposite circular polarization measure the population in K or K' valley respectively. (b) Illustration of valley relaxation of initially excited holes by exchanging spin and momentum with an electron in the other valley. (c) Illustration of valley relaxation of holes by scattering through the Γ valley. (d)Transient absorption spectra of monolayer $MoS_2$ at 74K with delay time varying from 0 ps to 15 ps between pump and probe beams. The red lines show SCP and blue lines show OCP respectively. The photo-induced absorption (PIE), stimulated emission (SA) and ground state bleaching (GSB) peaks are labeled in the figure. (e)Optical selectivity (η) decays with disorder (W). Lines with different colors represent different optical transition energies off the A exciton resonance of 1.860 eV. The numbers near lines show the transition energy relative to the direct band gap. (f) Schematic illustration of A and B excitonic transitions in the K and K' valley. The gray arrows show the exciton spin orientations. Coulomb-induced inter-valley coupling in TMDC materials leads to a decay of the optically generated valley polarization. The impurity assisted exchange term X leads to a carrier transfer between the resonant states in the two valleys (A–A,

B–B coupling). The Coulomb-induced renormalization term Y couples A and B excitons with the same spin at different valleys. It does not cause carrier exchange between the valleys but leads to energetic shifts of the excitonic states in both valleys. Panels (a) (b) (c) and (d) are reproduced from Ref. [28]; Panels (e) is reproduced from Ref. [26]; Panels (f) is reproduced from Ref. [31].

### 3.2.1. Bandgap Renormalization

With the presence of optical excited photo carriers in semiconductors, band gaps are "renormalized" by many-body effects arising from the presence of free carriers in the system due to exchange correlation corrections. Normally, this effect reduces the band gap and is usually referred to as the "bandgap renormalization" effect. Due to the enhanced many-body coulomb interaction, a result of extreme quantum confinement and reduced screening, the band gap renormalization effect is very prominent in thin layer TMDs. Eva A.A. Pogna et al. studied the bandgap renormalization effect with pump excitation density ($2-3*10^{12}$ cm$^{-2}$) which is below the Mott density in $MoS_2$.[55] In this intermediate density regime below the Mott-transition, the reduction of the band gap and the decrease of the binding energy tend to offset one with another as shown in Fig. 5(a), so the shift of main resonance of exciton is not significant. However, the compensation is partial and gives rise to an overall peak shift of a few tens millielectronvolts. Eva A.A. Pogna et al. use three different pump photon energies to excite a monolayer $MoS_2$ and probe the pump induced absorption of the sample with a broadband white light. As shown in Fig. 5(b), the transient absorption (TA) spectra recorded at fixed pump-probe delay t=300 fs exhibits three prominent features in correspondence of A, B, C exciton of $MoS_2$. Regardless of the pump photon energy, each feature of TA spectra consists of a reduced absorption at the excitonic resonance (negative signal) and a corresponding red-shifted photo induced absorption (positive signal). The simultaneous bleaching of the three excitonic transitions cannot be interpreted with Pauli blocking only, as in Pauli blocking, the transient change mainly happens around the exciton resonance that is closed to the pump excitation phonon energy. Time-domain ab-initio simulations including the effect of the renormalization of electronic gap and excitonic binding energy as shown in Fig. 5(c) indicate much better agreement with experiments. As shown in the schematic of Fig. 5(a) the overall shift of the excitonic absorption peak provides the alternating positive and negative peaks in correspondence of the excitonic resonances as observed experimentally.

With higher pump excitation density, the semiconductor undergoes a Mott transition from an insulating excitonic regime to an electron–hole plasma as shown in Fig. 5(d). In this region, the Coulomb interaction between charge carriers is strongly screened due to the high density of carriers, which reduces the exciton binding energy to zero. In this case, the exciton is no longer a bound state and the exciton resonance disappears. Alexey Chernikov et al studied the bandgap renormalization effect in mono/bi-layer $WS_2$ with intense optical pump pulses excitation (~$1*10^{14}$/cm$^{-2}$) lying in the Mott regime.[32] After strong pump excitations, the resonant exciton absorption features associated with the exciton resonance completely disappear in the optical response of bilayer $WS_2$ as shown in Figs. 5(e), and 5(g). At the same time, the reduction of the repulsive Coulomb interaction between charges of the same sign results in a decrease in the quasi-particle energy, which produces a giant bandgap renormalization, shifting the bandgap even below the energy of the initial exciton peak as shown in Fig. 5(f). The bandgap renormalization together with the filling effect of photo-excited electrons and holes contributes to the bleaching of the exciton resonance via Pauli blocking and building up of carrier population inversion near the band edge. This is magnified as a broad dip in the spectrum shown in Figs 5(e) and 5(g), several hundreds of meV below the initial excitonic transition, with an onset at 1.65 eV corresponding to renormalized band edge and a spectral width of 100 meV corresponding to population inversion in the electron and hole bands.

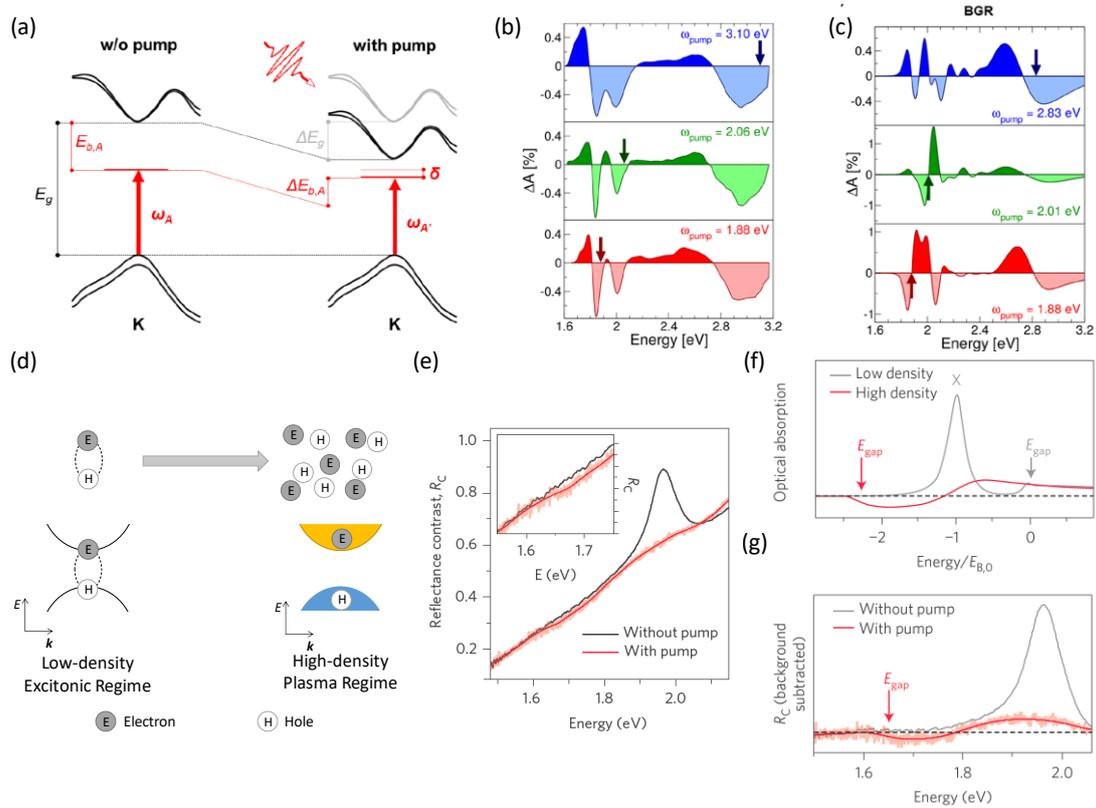

**Fig. 5.** (a) Band structure at K point in absence (left) and in presence (right) of pump pulse absorption. The red lines mark the optical excitation. Due to the presence of photo excited carriers, there is reduction of bandgap ($\Delta E_g$) and exciton binding energy ($\Delta E_b$), the overall shift of the excitonic absorption resonances: $\delta = \Delta E_g - \Delta E_b$. (b) Transient absorption spectra of 1L-$MoS_2$, recorded at fixed pump-probe delay t=300 fs for three pump photon energies that are in resonance with A (1.88 eV) and B (2.06 eV) exciton and out of resonance with C exciton (3.1 eV) respectively. (c) Calculated transient absorption spectra of 1L-$MoS_2$ for three pump photo energies by including renormalization of electronic gap and excitonic binding energy. (d) Transition from the low density excitonic regime to a dense electron-hole plasma. (e) Reflectance contrast spectra of $WS_2$ bilayer without pump excitation and 0.4 ps after excitation by a pump pulse at 70 K. (f) Schematic representation of the optical absorption at low and high carrier density regimes as predicted by theory. (g) Subtraction of a linear background from Fig. 5(d). Panels (a) (b) and (c) are reproduced from Ref. [55]; Panels (e) (f) and (g) are reproduced from Ref. [32].

### 3.2.2. Valley Selective Optical Stark Effect

The stark effect is the shifting and splitting of spectral lines of atoms and molecules due to presence of an external electric field. In optical stark effect, such external electric field is provided by the AC electric field of light. For exciton state in monolayer TMDs, the ground state and the exciton state can be approximated as a two level system similar to that in an atomic system as shown in Fig. 6(a). With the excitation of optical field with time average of electric field square $<\varepsilon_0^2/2>$, the corresponding energy shift is:

$$\Delta E = \frac{M_{ab}^2 <\varepsilon_0^2/2>}{\Delta}$$

where $M_{ab}$ is the polarization matrix element between the two states[33]. This interaction between the two states always results in a wider energy level separation, and the magnitude of the energy repulsion becomes more substantial if the two energy states ($\Delta$) are closer to each other. Furthermore, due to the distinct optical selection rules, the optical transitions of A excitons of K and K' valley couple to photons of opposite helicity: A exciton at K valley couples exclusively to left-circularly polarized light, and the A exciton at the K' valley couples only to the right-circularly polarized light, this not only governs resonant, but also non-resonant excitations. As a result, with circularly polarized pump excitation, taken left circularly polarized pump excitation as example, only A exciton in K' valley is coupled to the excited light field and the transition in K' valley is substantially modified through the optical stark shift, while leaving the exciton in K valley unchanged.

Using non-resonant circularly polarized pump excitation and broad band probe spectroscopy, Jonghwan Kim et al. and and Edbert J. Sie et al. have studied the valley selective stark effect in monolayer $WSe_2$[34] and $WS_2$ [33] respectively. As shown in Fig. 6(b), due to exciton resonance shift of stark effect, with pump excitation, the absorption spectrum of the probe will have a blue shift, this corresponds to a negative differential absorption ($\Delta\alpha$) on the blue side of the resonance and a positive differential absorption ($\Delta\alpha$) on the red side of the resonance as shown in Fig. 6(c). Although both resonant and non-resonant excitation can induce the stark effect, for experimental demonstration of this effect, non-resonant excitation is employed to avoid dissipation and dephasing naturally accompanying real excitations. Furthermore, if the pump excitation is circularly polarized, the stark effect is induced in one valley only, so the transient absorption change can be merely observed with probe light of the same circular optical helicity as shown in Fig. 6(d): a 4 meV blue shift of the K valley A-exciton resonance is induced in this case. This dispersive line shape cannot be due to the band renormalization effect described in the previous session, as non-resonant pump excitation is used in this case.

By applying intense circularly polarized light, which breaks time-reversal symmetry, the exciton level in each valley can be selectively tuned through the valley selective optical stark effect, this provides a convenient route to lift the valley degeneracy. On the other hand, valley degeneracy can be lifted by applying an external magnetic field through Zeeman effect[35-37], however, with 7 Tesla external magnetic field, the energy splitting is less than 1 meV in a monolayer WSe$_2$[36]. The photo induced valley energy splitting can be as large as 10 meV in WSe$_2$ through valley selective stark effect, which is equivalent to pseudo-magnetic field of 60 T.

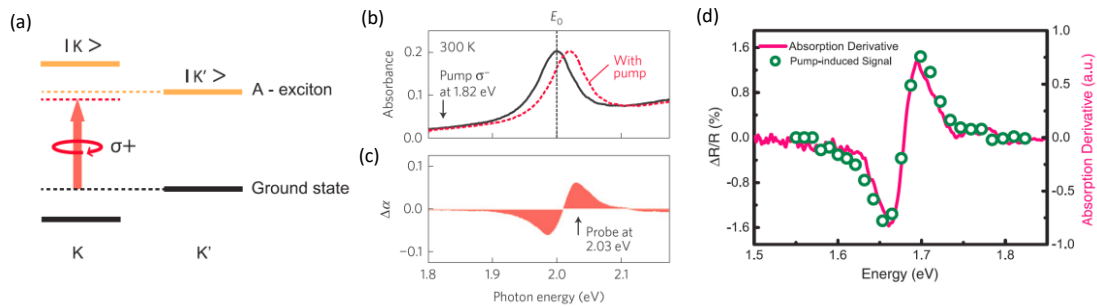

**Fig. 6.** (a) Schematic of the optical stark effect. The dashed black and yellow lines denote the unperturbed ground and exciton states, respectively. The pump photon energy is lower than the exciton resonance energy. (b)Absorbance of monolayer WS$_2$ with (red dash) and without (black) pump excitation that simulates the stark effect. (c) The simulated change of absorption induced by the pump pulses. (d) The transient reflection spectrum of monolayer WSe$_2$ at zero delay of probe. Panels (a) and (d) are reproduced from Ref. [34]; Panels (b) and (c) are reproduced from Ref. [33].

### 3.3. Time Resolved Kerr Measurement

Time resolved Kerr/Faraday rotation (TRKR/TRFR) is sensitive to the spin states of both photo created and resident carriers polarized by left or right circularly polarized pump laser. In a TRKR/TRFR measurements as shown in Fig. 7(a), a circularly polarized pump pulse generates valley-polarized excitons in a specific valley, the temporal evolution of the population imbalance of carriers with different spins can be probed by the polarization rotation of the reflected (Kerr) /transmitted (Faraday) probe pulse with a time delay t.

Due to the valley selective excitation with circularly polarized pump and spin valley coupling in TMDs, the sign of Kerr rotation would reverse if the helicity of the pump pulse is reversed as shown in Fig. 7(b). Early works by C. R. Zhu et al.[38] and S. Dal Conte et al[39] observe exciton valley depolarization time on the order of a few ps in WSe$_2$ and Mo(W)S$_2$ respectively using pump photon at or above the A exciton resonance. In their works, no evidence of transfer of spin/valley polarization to resident carriers in TMDs are observed. In the work by R.Z. Zhu et al., measurement on WSe$_2$ exhibits a mono-exponential decay of 6 ps at 4K as shown in Fig. 7(c), intervalley electron-hole Coulomb exchange interaction in bright exciton is considered as the dominating mechanism that accounts for the valley depolarization[40], and the exciton valley depolarization time decreases significantly as lattice temperature increases; on the other hand, the decay time changes little with excitation density, which means the exciton-exciton interaction probably play a minor role in the depolarization within this excitation density range (0.15~1*10$^{12}$/cm$^2$). In the parallel work by Conte et al[39], measurement on MoS$_2$ exhibits a bi-exponential decay as shown in Fig. 7(d): a fast component of 200 fs and a slow component of 5 ps at 77K. The slow component is also attributed to the intervalley electron-hole Coulomb exchange interaction, and the rapid component is attributed to the intervalley scattering of photo-excited excitons. In addition, this work has studied the excitation density dependence on a larger range (0.5~6*$10^{13}$/$cm^2$) and observed both the slow and fast component decreases with excitation densities.

Using probe photon energy below A exciton resonance, long lived transfer of spin/valley polarization to resident electrons and holes has been observed in doped Mo(W)$S_2$ and $WSe_2$ respectively with TRKR measurements[41, 42]. As shown in Fig. 7(e), taking n-doped case in $MoS_2$ for example, following circularly polarized pump excitation and initial recombination with photo generated holes, the spin distribution of resident electron densities maybe unequal and out of equilibrium. Thus, photoexcitation can impart a net spin polarization ($S_z = n_\uparrow - n_\downarrow + n'_\uparrow - n'_\downarrow$) and/or valley polarization ($N_v = n_\uparrow + n_\downarrow - n'_\uparrow - n'_\downarrow$) onto the resident electrons that may remain even after all holes have recombined. As in a TRKR/TRFR measurement, the rotation of the probe polarization only depends on the difference between a materials right and left circularly polarized (RCP and LCP) light absorption and indices of refraction. According to the selection rules of TMDs, RCP near the lowest energy A exciton in TMDCs couples to the resident electron density $n_\uparrow$ in the K valley (similarly LCP light couples to $n'_\downarrow$ in the K valley). Perturbations to the densities and shift the chemical potential which changes the absorption and refraction of RCP and LCP probe light at wavelength near optical transitions. Thus, to leading order, the Kerr signals are proportional to $n_\uparrow - n'_\downarrow$, to which both spin and valley polarization can contribute: $(S_z + N_v)//2$.

Luyi Yang et al.[42] use TRKR to measure the n-doped $MoS_2$ and $WS_2$ with probe photon energy just below A exciton resonance, the results as shown in Fig. 7(f) indicate that the decay of the optically induced electron polarization is extremely long (~3 ns) at 5 K. The spin lifetime decreases dramatically as temperature increases, at 40K, it already reduces to <200 ps. This temperature dependence matches the interpretation that the spin relaxation is dominated by Elliot-Yafet spin relaxation processes due to electron-phonon scattering with long wavelength flexural phonons. Similarly long spin life time (~ 1 ns) of resident holes/positive trion has been observed in $WSe_2$ by Wei-Ting Hsu et. al.,[41] as shown in Fig. 7(g). In both cases, the resident electron/hole spin lifetimes are 2-3 orders longer than typical radiative recombination time of exciton. The long lived Kerr signal can be explained by the transfer of valley pseudospin from photocarriers to the resident carriers. The transfer of the valley pseudospin can be mediated by trion, leaving behind valley polarized electron/hole after trion recombination. Wei-Ting Hsu et al.[41] further indicate that the long life time of resident hole can be observed only when pumping at the trion resonance instead of the neutral exciton resonance by perform pump photon energy dependent measurement as shown in Figs. 7(g) and 7(h): when excited with photon energy away from the trion resonance, the TRKR signal decreases dramatically, and the depolarization time also decreases to few ps which matches the exciton depolarization time observed by early TRKR works.[38, 39]

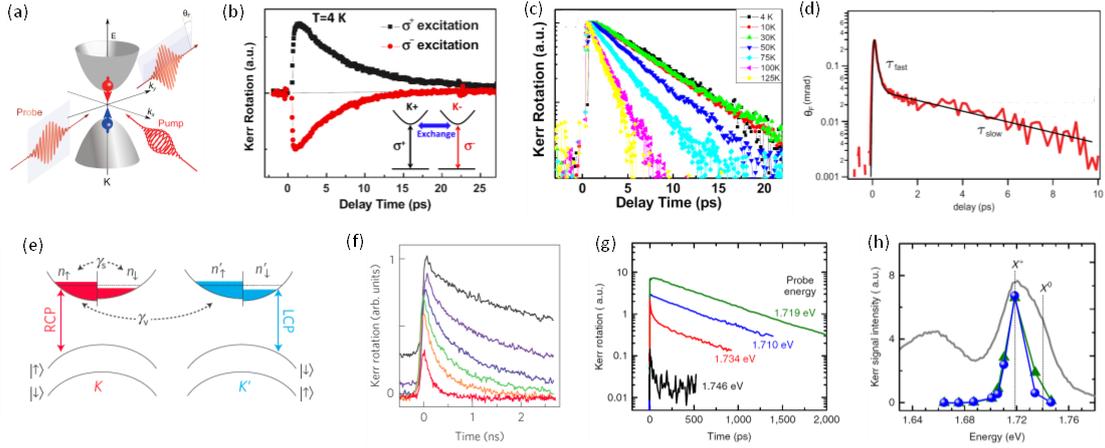

**Fig. 7.** (a)Schematic diagram of the TRFR experiment. The red and blue arrows represent the spin of the photo-excited electrons and holes. (b)Kerr rotation dynamics at 4 K for a σ+ and σ− pump beam. The laser excitation energy is 1.735 eV. The inset indicates optical selection rules of the excitons in K ± valleys and their coupling induced by the long-range exchange interaction. (c)Temperature dependence of Kerr rotation dynamics after a σ+ polarized pump pulse excitations. (d)Temporal evolution of θ$_F$(t) in semi logarithmic scale measured at 77 K. The black line is a bi-exponential fit to the data (red line). (e)Schematic illustration of a simple single-electron picture of the conduction and valence bands at the K and K' valleys of MoS$_2$ together with the relevant optical selection rules and scattering processes. The conduction bands are separately drawn as spin-up (K) and spin-down (K') components with slightly different curvature and small splitting. The n$_↑$, n$_↓$, n'$_↑$ and n'$_↓$ are densities of the resident electrons, and γ$_s$ is electron spin relaxation and γ$_v$ is spin-conserving inter-valley scattering. (f) Temperature dependence (red for 5 K and black for 40 K) of time-resolved Kerr rotation dynamics with pump pulse at 635 nm and 100 ps duration, and probe pulse at 672 nm and 250 fs duration. (g)Probe energy dependence of time-resolved Kerr rotation dynamics, the pump energy is kept at ~12 meV higher than probe photon energy. (h)The sketch of Kerr signal intensities at t=2.5 ps (green, solid triangles) and t=100ps (blue, solid circles) with the PL spectrum (grey curves) for comparison. The emission peaks of neutral exciton (X$^0$) and positive trion (X$^+$) are indicated by vertical dotted lines. Panels (a) and (d) are reproduced from Ref. [39]; Panels (b) and (c) are reproduced from Ref. [38]; Panels (e) and (f) are reproduced from Ref. [42]; Panels (g) and (h) are reproduced from Ref. [41].

## 4. Conclusion and Perspectives

In this review paper, we have discussed most, if not all, helicity resolved ultrafast spectroscopy measurements that concern monolayer TMDs so far and try to draw a consistent conclusion regarding the spin/valley dynamics from a comprehensive consideration of these results obtained by different ultrafast spectroscopy measurements. Due to the extremely large many-body Coulomb interactions in atomically thin TMDs, band renormalization dominates the transient absorption response with above optical gap pump excitation of photo carriers. The Coulomb induced bandgap renormalization together with impurity induced intervalley carrier transfer also account for the prominent spectral features in both valley immediate after the pump excitations. While with pump excitation below bandgap, the optical stark effect dominates the response and if excited with circularly polarized light, the optical stark effect induce a very large pseudo magnetic field (~60 Tesla) and breaks the valley degeneracy.

The radiative lifetime of exciton and trion are relatively short (~ ps for exciton and over 10 ps for trion) at low temperature. Generally, the trion radiative lifetime is longer than exciton because the oscillator strength of trion is lower due to a stronger localization comparing to exciton. As radiative lifetimes of exciton and trion are both relatively short, the intervalley scattering process is generally not observed in a PL measurement of exciton state; for trion state, the fast component of intervalley scattering time is still short (~10 ps), due to the efficient intervalley exchange interaction. However, the valley pseudospin will transfer from photo-excited valley polarized trion to the resident carriers, leaving behind valley polarized electron/hole, this would develop a much longer spin polarization life time after the radiative recombination as a single particle spin slip is needed for the residue electron/hole which usually takes longer. For this case, spin life time up to 3 ns (1 ns) is observed for electron (hole), which is 2-3 orders larger than typical radiative recombination time of exciton states.

In this work, we only concern the valley/ spin dynamics in single species of TMDs, which provides the starting point and building block towards the full understanding and manipulation of valley degree of freedom for various possible valley based applications. For future work, helicity resolved ultrafast spectroscopy has to be applied on TMDs with better control of various material parameters such as defects, magnetic dopants, both for better understanding of valley/spin dynamics and for better quantum control of the valley degree of freedom. For multi-functional devices, van der Waals heterostructures[64] based on TMDs provides infinite opportunities, but their valley/spin related dynamics remain to be explored experimentally, especially with ultrafast spectroscopy[65].

**Acknowledgment**

The authors would like to acknowledge the funding support from the following: the National Basic Research Program of China (973 Grant Nos. 2012CB921300, 2014CB920900), the National Natural Science Foundation of China (NSFC Grant No. 11274015, 11674013, 21405109), National Key Research and Development Program of China (Grant No: 2016YFA0300802), the Recruitment Program of Global Experts and Beijing Natural Science Foundation (Grant No. 4142024).